# Comparative Study of Simulation for Vehicular Ad-hoc Network

Prof. Vaishali D. Khairnar
Symbiosis Institute of Technology
Lavale Pune - Maharashtra
India.

Dr. S. N. Pradhan
Nirma Institute of Technology
Gandinagar- Ahmedabad
India

## ABSTRACT

In this paper we have discussed about the number of automobiles that has been increased on the road in the past few years. Due to high density of vehicles, the potential threats and road accident is increasing. Wireless technology is aiming to equip technology in vehicles to reduce these factors by sending messages to each other.

The vehicular safety application should be thoroughly tested before it is deployed in a real world to use. Simulator tool has been preferred over out door experiment because it simple, easy and cheap. VANET requires that a traffic and network simulator should be used together to perform this test. Many tools exist for this purpose but most of them have the problem with the proper interaction. Simulating vehicular networks with external stimulus to analyze its effect on wireless communication but to do this job a good simulator is also needed.

## 1. INTRODUCTION

Traffic congestion on the roads is today a large problem in big cities. The congestion and related vehicle accommodation problem is accompanied by a constant threat of accidents as well. According to National Highway Traffic Safety Administration, the following figures indicate some of the consequences of recent car accidents.[1]

• Police reported traffic accidents

• 400 people were killed

• 400 of people were injured

**Table 1:-** Motor Accidents in the past 8 years.

| Year | Fatal | | Serious | | Minor | | Without-Injury | Total No.Of Accident |
|---|---|---|---|---|---|---|---|---|
| | No.Accidents | No. of Dead | No. of Accidents | No. of Injured | No. of Accidents | No. of Injured | | |
| 2001 | 260 | 262 | 210 | 241 | 1470 | 1013 | 242 | |
| 2002 | 293 | 309 | 186 | 201 | 1331 | 1714 | 171 | |
| 2003 | 305 | 324 | 168 | 214 | 1360 | 1478 | 94 | 1927 |
| 2004 | 348 | 364 | 270 | 303 | 1383 | 1529 | 112 | 2113 |
| 2005 | 334 | 344 | 284 | 209 | 1215 | 1342 | 146 | 1979 |
| 2006 | 360 | 372 | 309 | 338 | 1257 | 1408 | 197 | 2123 |
| 2007 | 403 | 414 | 406 | 462 | 1352 | 1575 | 186 | 2347 |
| Up To 30 Sep 2008 | 303 | 343 | 413 | 487 | 792 | 937 | 153 | 1688 |

**Table 2:-** Accidents Occurred in Pune City from 2003 to 30 September 2008

| Sr.no | Year | Fatal | | Serious | | Minor | | Nor-Inj.Accidents | Tota No. Of Accidents |
|---|---|---|---|---|---|---|---|---|---|
| | | Accident | Dead | Accident | Injured | Accident | Injured | | |
| 1 | 2003 | 300 | 324 | 168 | 216 | 1360 | 1478 | 94 | 1927 |
| 2 | 2004 | 348 | 364 | 270 | 303 | 1383 | 1529 | 112 | 2113 |
| 3 | 2005 | 334 | 344 | 284 | 209 | 1215 | 1342 | 146 | 1979 |
| 4 | 2006 | 360 | 372 | 309 | 338 | 1257 | 1575 | 197 | 2347 |
| 5 | 2007 | 403 | 414 | 406 | 462 | 1352 | 1575 | 186 | 2347 |
| 6 | upto 30 Sep 2008 | 330 | 343 | 413 | 487 | 792 | 907 | 153 | 1688 |

Preliminary precautions like seat belts and airbags are used but they cannot eliminate problems due to driver's inability to foresee the situation ahead of time. On a highway a vehicle cannot currently predict the speed of other vehicles. However, with use of wireless sensor, computer and wireless communication equipment, speed could be predicted and a warning message may be sent every 0.5 seconds could limit the risk of potential accidents. Wireless communication is can be applied to different scenarios. Vehicular Ad Hoc Networks (VANETS) is one of its types which deploys the concept of continuously varying vehicular motion. The nodes or vehicles as in VANETS can move around with no boundaries on their direction and speed. This arbitrary motion of vehicles poses new challenges to researchers in terms of designing a protocol set more specifically for VANETS. Tests are being carried out through simulated environments to check the way VANETS perform, before they are used in commercial application in the real world.

## 2. Features of VANET

• The nodes in a VANET are vehicles and road side units

- The movement of these nodes is very fast

- The motion patterns are restricted by road topology

- Vehicle acts as transceiver i.e. sending and receiving at the same time while creating a highly dynamic network, which is continuously changing.

- The vehicular density varies from time to time for instance their density might increase during peak office hours and decrease at night times.

## 3. VANET Simulation Problem

To evaluate VANET protocols and services, the first step is to perform an outdoor experiment. Many wireless technologies such as GPRS, IEEE 802.11p and IEEE 802.16 have been proposed for reliable traffic information. For test purpose software simulations can play a major role in imitating real world scenarios.

VANET relies on and is related to two other simulations for its smooth functioning, namely traffic simulation and network simulation. Network simulators are used to evaluate network protocols and application in a variety of conditions. The traffic simulators are used for transportation and traffic engineering. These simulations work independently but to satisfy the need of VANET, a solution is required to use these simulators together. There are a large number of traffic and network simulator and they need to be used together into what can be called VANET simulator. There are few tools for VANET simulation but most of them have the problem of proper 'interaction'. Therefore, comparison of present tools will suggest the choice with proper 'interaction'. This paper consists of a survey of various traffic simulators, network simulators and VANET simulators resulting in the selection of a preferred recommended choice.

### 1) GrooveSim

GrooveSim [16] was the first tool created for evaluation of VANET performance mainly motivated by vehicular traffic flow and forecasting. The concept of application involves testing the possibilities of real time events as time-critical safety messages. GrooveSim was coded in C++ and Matlab provides GUI for drawing structures and graphs.

GrooveSim could operate in five different modes: predetermined, on-road, simulation, hybrid, research. A group of five vehicles travelling around the city and highway were simulated for recording certain parameters like message penetration, delay, vehicle grouping, packets dropped etc and packet time to live values were calculated in the simulator. GrooveSim did not include any network simulator and also it was unable to create traces for any network simulator.

### 2) NHTSA (National Highway Traffic Safety Application)

NHTSA (National Highway Traffic Safety Application) [13] provided VANET estimation and focused on a global perception of VANET performance. This platform is a computer-based tool and accepts a text file during vehicular simulation. The NHTSA simulator was designed for networking research and was built on the top of NS-2 simulator. The simulator is platform-independent and is capable to run on both Win32 as well as Linux. It has strong GUI support implemented by C++. The main purpose of NHTSA project was to promote DSRC standardization, and during the test-bed, a GPS receiver, Windows-based notebook and IEEE 802.11a wireless device are used as a hardware module for DSRC standard. The platform is very scalable and flexible for researchers to alter the configuration according to the requirements.

### 3) FleetNet

Aim is to provide a platform based on simulation results from simulation tools and a software prototype called FleetNet Demonstrator FND[12,17]. The development of this software was aimed to state the problems found in inter-vehicular communication and realistic evaluation of VANET. The focus of was primarily on how mobility can be achieved with position based routing protocols. The demonstrator of the project performed an outdoor experiment with six vehicles. Each vehicle had two laptops, one as a Linux system for the communication between the vehicles and vehicular to infrastructure through WIFI card. The other laptop had a windows system to provide a GUI for vehicular communication as well as communication with the GPS receiver. The demonstrator concluded some results by inspecting the vehicular behavior on highways and in the city, the transmission of data, velocity and distances amongst vehicles.

### 4) CARLINK

The CARLINK [14] was developed to provide a wireless traffic service platform between the cars. Vehicles were equipped with wireless transceivers to communicate with road-side infrastructure. Vehicles were also able to form ad-hoc network with each other. The base station was able to collect car real-time data such as local weather, traffic density and all information about current traffic and pass them to central unit for database updating, which is then sent back to the vehicles driving past the base-station.

The CARLINK project developed applications like FSF (Finding and Sharing Files) and PB (Ad hoc puzzle bubble) [15] on the java based application called JANE. The purposes of the applications were to facilitate researchers to apply their tests in simulated environment before being installed into real ad-hoc network. Both these applications could be installed into the laptops and PDA.

### 5) Car2Car

The Car2Car[3] communication group is an organization instigated by European vehicle manufacturers that is open for providers, research associations and other partners. Car2Car uses IEEE 802.11 WLAN technology and a frequency spectrum in the 5.9 GHz range has been allocated on a harmonized basis in Europe in line with similar allocations in USA. As soon as two or more vehicles or ITS stations are in radio communication range, they connect automatically and establish an ad hoc network where all ITS stations know the position, speed and direction of the other stations and will be able to provide warnings and information.

As the range of a single Wireless LAN link is limited to a few hundred meters, every vehicle is also router and allows sending messages over multi-hop to farther vehicles and ITS stations. The routing algorithm is based on the position of the vehicles and is able to handle fast changes of the ad hoc network topology for the vehicles to correspond with each other within the range of hundred meters and forms an ad hoc network. The routing algorithm verifies the location and speed of a vehicle and is able to oppose changing in the topology if any. Car2Car communication is based on the following points.

• Advance Driver Assistance. Design and development of active safety applications

• Decentralized Floating Car Data

• User Communication and Information Services

## 6) Clarion

The Clarion[4] Communication method for Intelligent Multimode Transit System (IMTS) utilizes a spread spectrum for the automatic control of multiple vehicles. Information such as the speed and position of the vehicles is transmitted by the transceivers installed on each vehicle. Clarion is a low cost budget solution. At present this technology is set up in the amusement park which controls the shuttle bus service within the park.

## 7) IP PReVENT

PReVENT[8] is a EU sponsored project to demonstrate safety application using sensors, maps and communication systems. PReVENT will contain 23 cars, trucks and different types of simulators for active safety including,

• Safe speed and safe following

• Collision Migration

• Intersection safety

• Lateral Support

• Development of ADAS (Advance driver assistance system)

• Using maps for improved ADAS

• Evaluation of ADAS

• Sensor data fusion

## 8) Cooperative Vehicles and Infrastructure Systems (CVIS)

CVIS is a developed for the purpose to increase road safety and effectiveness and reduce the environmental impact of road safety. CVIS tests technologies to permit vehicles to communicate with each other and near by road side.

**The main objective is[9] :**
• Development of standards for the vehicle-to-vehicle and vehicle-to-vehicle infrastructure communication.

• Bringing more precision in the vehicle location and generation of more dynamic and accurate local maps using satellite navigation and other modern methods of location referencing.

• New systems for cooperative traffic and network monitoring in both vehicle and roadside infrastructure and to detect incidents immediately.

• Range of cooperative applications for traffic management, mobility services and driver assistance.

• Development of toolkits to address key deployment.

Local floating Car data application: The service updates the service centre about different parameters of vehicles.

## 9) Demo 2000

This is the joined together project by Tsukuba and Japan for the demonstration of the cooperative driver assistance system DEMO 2000[6].Aim to evaluate feasibility and technologies for inter vehicle communication and are linked together to communicate with each other. DOLPHIN (Dedicated Omni Purposed Inter-vehicle linkage) protocol was used in 5.8 GHz DCRC and CSMA was used to access medium. Each vehicle was equipped with laser radar for the measurement of distance, obstacles, and LCD for displaying vehicle communication.

## 10) Car Talk 2000

This is European project to help driver based on communication between vehicles, which consists of self ad-hoc radio network with the purpose of developing future technology. Car talk 2000 [5,10] provides reliable components for Advance driver Assistance (ADAS) such as Advance cruise control (ACC). With the different approaches, the communication can greatly improve and provide better safety and fewer injuries that are caused by collisions. The objectives of car talk 2000 are safety and to evaluate the advance driver assistance.

car talk 2000 creates application in the following way:

• Information and warning function.

• Communication based longitudinal control system.

• Co-operative assistance system.

**Information and warning function:**
All the parameters are informed by the mean of information signal. These parameters include traffic load, conditions, road accidents etc. This prior information allows a driver a pre-sense capability about safety measures.

**Communication-based longitudinal control system:**
The ACC systems were only capable of capturing about the vehicle in the front but it is now possible to know about any vehicle in front that is braking. This leads to more favourable behaviour and the avoidance of any collision that is due to the vehicle in the front.

**Co-operative Assistance System:**
As nowadays, misunderstanding between the drivers on merging points occurs. It will help us in merging points. The speed and lane of the entire vehicle must be known in advance for better communication.

## 11) Comparison of Simulators

VANET technology enables communication between vehicles and nearby road-side infrastructure [18] and is made possible through a wireless sensing device installed in the vehicles. With the inception of VANET, new opportunities and related technologies like applications for traffic jam, accident control and weather updates have appeared.

VANET performance can be tested in real situations but factors like cost, inaccurate results and protocol evaluation of complex environment may contribute towards a disappointing end.

| Simulator | GloMoSim | Ns-2 | NCTUns | | QualNet |
|---|---|---|---|---|---|
| Signal to Noise Ratio Calculation | Cumulative | Difference in two Signals | Cumulative | | Cumulative |
| Signal Reception | SNRT, BER | SNRT | Sender: Transmitting power | Receiver: Power threshold, Distance | SNRT, BER |
| Fading | Rayleigh, Riecan | No | Rayleigh, Riecan | | Rayleigh, Riecan |
| Path Loss | Free Space, Two Ray | Free Space, Two Ray | Free Space, Two ray, Free space with shadowing | | Free Space, Two Ray, ITM (Irregular Terrain Model) |
| Support for Multiple Wireless Technology | Yes | No | Yes | | Yes |
| Antenna's Support | Bi-directional, Omni-directional | Bi-directional, Omni-directional | Directional, Bi-directional, Rotating | | Bi-directional, Omni-directional, beam, Switched |

| | GloMoSim | Ns-2 | NCTUns | QualNet |
|---|---|---|---|---|
| | | | | beam |
| Distributed Simulation | Yes | No | Yes | Yes |
| Time required for Simulating 5000 Nodes (sec) | 6191 | Fail | Fail | 6191 |
| Memory Required for Simulating 5000 Nodes (KB) | 27.5 | Fail | Fail | 27.5 |
| GUI | Yes | No | Yes | Yes |

## 12) Conclusion

I have explored just the initial developments that were carried out in creating a simulator which can be aimed will testing VANET.

I would like to acknowledge my Guide Dr. S. N. Pradhan for his guidelines and support. Also I would like to acknowledge my Husband for valuable support.